\documentclass[12pt]{article}
\usepackage[dvips]{color}
\usepackage{epsfig}
\usepackage{amsmath}
\usepackage{graphicx}

\date{}
\begin{document}
\title{\bf Logarithmic correction of the BTZ black hole and adaptive model of Graphene}
\author{{Behnam Pourhassan$^{a}$\thanks{Email: b.pourhassan@du.ac.ir},
Mir Faizal$^{b,c}$\thanks{Email: mirfaizalmir@googlemail.com}, and S. Ahmad Ketabi$^{a}$\thanks{Email: saketabi@du.ac.ir}}\\
$^{a}${\small {\em School of Physics, Damghan University, Damghan, 3671641167, Iran}}\\
$^{b}${\small {\em Deptartment of Physics and Astronomy, University of Lethbridge,}}\\
{\small {\em Lethbridge, AB T1K 3M4, Canada}}\\
$^{c}${\small {\em Irving K. Barber School of Arts and Sciences,  University of British Columbia - Okanagan,}}\\
{\small {\em Kelowna, BC V1V 1V7, Canada}}}
\maketitle
\begin{abstract}
It is known that almost all approaches to quantum gravity produce a logarithmic correction term to the entropy of a black hole,
but the exact coefficient of such a term varies between the different approach to quantum gravity. Such logarithmic terms can also occur
due to thermal fluctuations in both analogous and real black holes so that we will analyze the effects of
logarithmic corrections  term with variable coefficient on properties of  analogous black hole.
As these properties can be experimentally tested, they can be used
to obtain the correct coefficient for such terms for an analogous black hole.
We will argue that as even the real black holes can be considered as thermodynamical objects in Jacobson formalism,
so such analogous  black holes can be used to obtain the correct coefficient for the  real black holes, and this in turn can be used
to select the correct approach to quantum gravity. In that case, we use an adaptive model of graphene, which is still far from real graphene, to investigate some thermodynamics quantities of BTZ black hole.
\end{abstract}

\section{Introduction}
It is important to understand how gravity is quantized to understand physical systems like the black holes
and even the physics at the big bang. The problem with testing quantum gravity
is that the energy scale at which quantum gravity
operates is so large that it has not been
possible to directly test quantum gravitational effects.
Furthermore, there are various approaches to quantum gravity,
and they seem to take  different approaches to describe  the fundamental structure of space-time.
Thus, it becomes even more difficult to test quantum gravitational effects. However, there is
one result that seems to be almost a universal predication of all theories of quantum gravity.
This prediction is that the leading order term in the entropy of a black hole is proportional to the
area of the black hole, $S_0 \sim A$.
There are good reasons for such a prediction, as this result can also be
produced from semi-classical approximation. So, it cannot be a purely quantum gravitational
effect, but it seems to be an effect that is generated because of quantum field theory in a classical
space-time.
It may be noted that it is not possible to measure the black hole thermodynamics directly, and so it
is difficult to use the black hole thermodynamics to test any quantum gravitational effect.
However, it is possible to use analogous black hole like adaptive model of
graphene to analyze such systems, as it is possible to construct
an effective BTZ-like black hole geometry in graphene \cite{1, u}. In this effective BTZ-like solution, the
velocity of light gets replaced by the Fermi velocity
($v_{f}\approx0.003 c$), and other constants are also defined effectively. This is because the
effective field theory describing
a graphene sheet resembles a massless Dirac equation
in three dimensions \cite{5b}.
A sheet of   graphene   is made up of carbon atoms
  arranged in hexagonal structures. In this structure, there exists a
triangular lattice with two atoms per unite
cell. So, the   lattice vector for graphene is given by
\begin{eqnarray}
{\bf a}_1 = \frac{a}{2}\left(3, \sqrt{3}\right), &&
{\bf a}_2 = \frac{a}{2}\left(3, -\sqrt{3}\right),
\end{eqnarray}
The  carbon-carbon distance for this structure can be written as
\begin{eqnarray}
 a \sim 1.42 A.
\end{eqnarray}
The reciprocal lattice vector for graphene is given by
\begin{eqnarray}
{\bf b}_1 = \frac{2\pi}{3a}\left(1, \sqrt{3}\right), &&
{\bf b}_2 = \frac{2\pi}{3a}\left(1, -\sqrt{3}\right).
\end{eqnarray}
Furthermore,  two points at the corners of graphene Brillouin zone are called the Dirac points, and they
are located at
\begin{eqnarray}
K = \left(\frac{2\pi}{3a} , \frac{2\pi}{3\sqrt{3}a}\right), &&
K'  = \left(\frac{2\pi}{3a} , -\frac{2\pi}{3\sqrt{3}a}\right).
\end{eqnarray}
In this structure, the  three nearest neighbors can be represented by
\begin{eqnarray}
 \delta_1 =  \frac{a}{2}\left(1, \sqrt{3}\right), & \delta_2 = \frac{a}{2}\left(1, -\sqrt{3}\right),
 & \delta_3 = -a \left(1, 0\right).
\end{eqnarray}
The six next to the nearest neighbors can be represented by
\begin{eqnarray}
 \delta_1 =  \pm \bf{a}_1, & \delta_2 = \pm \bf{a}_2,
 & \delta_3 = \pm \left(\bf{a}_1 - \bf{a}_2\right).
\end{eqnarray}
 The electrons in this structure are described by  the tight-binding approach. So, electrons in graphene can hop to both
nearest atoms and next to the nearest atoms and the
 Hamiltonian for graphene can be written as
 \begin{eqnarray}
  H = -t \sum (a^\dagger_\sigma b_\sigma + h.c, )
  - t' \sum  ( a^\dagger_\sigma a_\sigma + b^\dagger_\sigma b_\sigma +
  h.c.),
 \end{eqnarray}
where $\sigma$ denotes the spin. The
 hopping energy to the nearest neighbor is   $t\sim 2.8 eV$, and the hopping energy to the next to the nearest
neighbor is denoted  by $t'$. So, the   energy bands for graphene can be written as \cite{tigh}
\begin{eqnarray}
 E_\pm = \pm t \sqrt{3  + f(k)} - t' f(k),
\end{eqnarray}
and here $f(k)$ is defined as
\begin{eqnarray}
 f(k)  = 2 \cos \sqrt{3} ( k_y a ) +
 4 \cos \left(\frac{\sqrt{3}}{2} k_y a \right)
 \cos \left(\frac{3}{2} k_x a\right).
\end{eqnarray}
If this energy band is  expanded around the Dirac point, then we will obtain  \cite{tigh}
\begin{eqnarray}
 E_\pm \sim  \pm v_f |q|,
\end{eqnarray}
where $k = K + q$, and a similar expression can be obtained  for $K'$. The  velocity
$v_{f}\approx0.003 c$ is   the Fermi velocity. Now, to obtain the
  Hamiltonian of graphene close to the Dirac point,  the following approximation can be used
\begin{eqnarray}
 \sum_i e^{\pm K \delta_i} = \sum_i e^{\pm K' \delta_i} =0.
\end{eqnarray}
It may be noted that near the Dirac point, in the continuum limit,
this Hamiltonian for graphene  is the resembles the  Dirac Hamiltonian
 \cite{tight}. The only difference between this Hamiltonian and the usual Dirac
 Hamiltonian is that the velocity of light is replaced by the Fermi velocity.
In fact, it has also been
verified experimentally that the graphene is described
by a massless Dirac equation in three dimensions \cite{6b,7b}.
So, the effective field theory of a flat sheet of  graphene
is the Dirac equation in $(2+1)d$   space-time, with an effective Lorentz symmetry.
However, in this effective field theory, the
velocity of light being replaced by Fermi velocity.
This is the   main difference between the relativistic  Dirac
equation in three dimensions
and this effective field theory describing graphene \cite{5b}.

It is also known an effective curvature like effects can be induced in a deformed
sheet of  graphene. A graphene sheet has been used for
constructing the surface of revolution with constant negative curvature \cite{1a}.
A gravito-magnetic
field can be  introduced because of rotation, and its effects
on particles  resemble that of a  magnetic field. So,  an  externally
magnetic field   would induce
an effective gauge field   on the  graphene  sheet.
This would break time reversal invariance but not conformal invariance.
Thus, it has been argued that the curved
graphene sheet with  constant negative
curvature in the externally applied electromagnetic field could be modeled by
the stationary optical metric of the Zermelo
form, and it is conformal to the BTZ black hole \cite{1, u}.
Thus, a deformation of the graphene sheet can produce a BTZ  black hole like solution,
and in this BTZ   black hole
like solution,  the velocity of light is replaced by the Fermi velocity.
So, it is possible to analyze
an effective BTZ-like black hole solution in graphene \cite{1, u1}. It should have pointed out that it is only suggestive and the real graphene is still far.
It may be noted that the horizon of a usual BTZ black hole traps photons, but the
effective horizon of the black hole in graphene traps fermions moving with the Fermi velocities.
Entropy is associated with all black objects in  general relativity \cite{1q, 1aqp},
and so we expect that an entropy
should also be associated with the black hole like solutions in graphene.
In fact, it has been demonstrated that Hawking radiation is radiated from
such solution in graphene
\cite{u, u1, u2}. It may be noted that the Hawking radiation from
such a solution is an effective phenomenon,
and it occurs as the effective horizon traps Fermi velocities.

In this paper, we will use the effective black hole thermodynamics in the analogous
BTZ like black hole solution
in the adaptive model of graphene, which is still far from the real graphene,  to gain the  correct approach to quantum gravity.
This will be done
by analyzing the effect of the corrections to the entropy of such an effective
BTZ-like black hole solution in the adaptive model of graphene. Such quantum corrections of rotating BTZ black hole recently studied by the Ref. \cite{6666}.
\section{Logarithmic correction}
Almost all approaches to quantum gravity also predict that the leading order
correction to the entropy of
a black hole would be a logarithmic correction.
This logarithmic correction has been obtained using non-perturbative quantum  general
relativity  \cite{1z}.   In this approach, the relation between the
 density of states of a black hole and
the conformal blocks of a well-defined  conformal field theory have been used to obtain this
logarithmic correction.
The Cardy formula has also been used for obtaining such correction terms \cite{card}.
An exact partition function has been used for analyzing the correction to the entropy of a BTZ black hole,
and it has been demonstrated that the leading order corrections are to the entropy of a BTZ black hole  are
logarithmic corrections \cite{card}. The leading order corrections terms for dilatonic black holes have also been
demonstrated to be logarithmic corrections \cite{jy}. String theory has also been used for analyzing the corrections
to the entropy of a black hole, and it has been observed that string theoretical effects also produce a logarithmic
correction to the entropy of a black hole \cite{solo1, solo2, solo5}.  It has been demonstrated that the generalized
uncertainty principle has also  generated logarithmic correction terms
for the entropy of a black hole
\cite{mi, r1}.

Even thought the existence of such a logarithmic term  is predicted by almost all approaches to quantum gravity,
the exact coefficient of such a term is different in different approaches to quantum gravity. Thus, we can write the
general form for the leading order correction terms for the entropy of a black hole as
$ S =  S_0 + S_1$, where $S_0 \sim A$ and  $S_1 \sim \ln A $.
The  quantum fluctuations become important when the size of the black hole becomes sufficiently small. However, the temperature of a black
hole increases as the size of the black hole becomes small, and  so the effects of thermal fluctuations also increase as the size
of the black hole becomes small \cite{EPJC}. Hence, it seems that the  thermal fluctuations in the thermodynamics of a black hole can be related to the
quantum fluctuations in the geometry of a black hole.  This is more obvious in the Jacobson formalism \cite{z12j,jz12}. This is
because in the Jacobson formalism,
that the Einstein's equations are obtained the  thermodynamics  by requiring that the
Clausius relation holds for all the local Rindler causal horizons through each space-time  point \cite{z12j,jz12}.
So, in the Jacobson formalism, thermal fluctuations in the thermodynamics will
generate  quantum fluctuations in the geometry of the space-time. Corrections to the thermodynamics of
black holes from such  thermal fluctuations have been analyzed \cite{l1,SPR}, and it was observed that such corrections are again
logarithmic corrections.

In the Ref. \cite{S1} the black hole entropy corrections due to thermal fluctuations in the various AdS black hole extensive parameters have been studied. It was found universality in the logarithmic corrections to charged black hole entropy in various dimensions. These corrections are expressed regarding the black hole response coefficients via fluctuation moments. Also, the black hole entropy correction was calculated from the quantum correction point of view using the near horizon conformal algebra \cite{S2}.

 In fact, the entropy of a very small black hole after thermal fluctuations were taken into account can be written as
$S=S_0+ S_1 = S_{0}- \ln{S_{0}T^{2}}/ 2$. As the existence of a logarithmic term is a universal feature of all approaches to
quantum gravity, but the exact coefficient depends on the exact approach to quantum gravity chosen, we will not fix the  coefficient
of the logarithmic term, and write the corrected entropy of the black hole as $S=S_{0}-\alpha \ln{S_{0}T^{2}}/ 2,$ where
$\alpha$ is a parameter that depends on the details of the model used. It may be noted that for thermal fluctuations in the
Jacobson formalism \cite{z12j,jz12}, $\alpha =1$ holds for very small black holes where the effects of thermal fluctuation have to be
taken into consideration, and for very large black hole where the effects of thermal fluctuations can be neglected it is possible to take
$\alpha =0$.  The effect of
thermal fluctuations on the  thermodynamics of black holes in anti-de Sitter
space-time  has been  studying using this formalism \cite{adbc}.
This formalism has also been used for analyzing the corrections to the  thermodynamics of a black Saturn \cite{dabc,dabc1} and modified Hayward black hole \cite{dabc2}. In the Ref. \cite{111} we have been study P-V criticality of logarithm-corrected dyonic charged AdS black holes as well as AdS black holes in massive gravity \cite{222} and investigate thermal fluctuation effects. Also, thermodynamics of an infinitesimal singly spinning Kerr-AdS black hole investigated with mentioned logarithmic correction \cite{333}. Such logarithmic term considered for the STU black hole and found important modification in the thermodynamics and also hydrodynamics \cite{444}.Logarithmic correction as quantum effects considered recently for the Horava-Lifshitz black hole \cite{5555}.
As logarithmic correction terms occur in all approaches to quantum gravity,
we can analyze such effects using an adaptive model of graphene. In that case, already we study that using dumb holes \cite{555}.
Furthermore, as such terms can be generated from thermal fluctuations, such terms  will occur in the adaptive model of graphene.
\\\\

\section{Thermodynamics}
So, now we will analyze the thermodynamics of an effective BTZ-like solution in the adaptive model of graphene, with the velocity of light
effectively replaced by the Fermi velocity, and this  effective
BTZ-like  black holes  in adaptive model of graphene can be written as  \cite{1, u},
\begin{equation}\label{1}
ds_{BTZ}^{2}=\frac{dr^{2}}{\Delta}+r^{2}d\phi^{2}+\left(\frac{J^{2}}{4r^{2}}-\Delta\right)v_{f}^{2}dt^{2}-Jv_{f}dtd\phi,
\end{equation}
where  $\Delta= {r^{2}}({l^{2}})^{-1}-M+{J^{2}}({4r^{2}})^{-1}$, $M$ is mass and $J$ is the angular momentum of black hole.
The black hole horizons obtained by setting $\Delta=0$.
Thus, for a BTZ  black hole \cite{BTZ} like solution in graphene \cite{1, u1}, we can write
$
r_{\pm}^{2}= {l} (lM\pm\sqrt{(lM)^{2}-J^{2}})/2,
$
where $r_{+}$ is the outer horizon, and $r_{-}$ is the inner horizon. It may be noted that the
product of two horizons is independent of mass,
$
r_{+}r_{-}= {lJ}/{2}
$. The standard entropy of the black hole is given by
\begin{equation}\label{5}
S_{0\pm}=\frac{A_{\pm}}{4}v_{f}^{3}=4\pi r_{\pm}v_{f}^{3}=2\sqrt{2}\pi \sqrt{l(lM\pm\sqrt{(lM)^{2}-J^{2}})}v_{f}^{3},
\end{equation}
where $S_{0\pm}$ is  original entropy, which is obtained by neglecting thermal fluctuations.
The Hawking temperature of the inner horizon and the outer  horizon is given by,
\begin{eqnarray}\label{6}
T_{\pm}&=&\frac{\kappa_{\pm}}{2\pi}=\frac{1}{4\pi}\left(\frac{2r_{\pm}}{l^{2}}-\frac{J^{2}}{2r_{\pm}^{3}}\right)\nonumber\\
&=&\frac{\sqrt{2}}{2\pi l^{\frac{3}{2}}} \frac{(lM)^2\pm lM\sqrt{(lM)^{2}-J^{2}}-J^{2}}{(lM\pm\sqrt{(lM)^{2}-J^{2}})^{\frac{3}{2}}}.
\end{eqnarray}
Here the surface gravity of the inner is denoted by  $\kappa_{-}$ and the surface gravity of the outer horizon is denoted by
 $\kappa_{+}$. It is possible to express the product of temperatures as
\begin{equation}\label{7}
4\pi^{2} T_{+}T_{-}=\frac{16r_{+}^{4}r_{-}^{4}-4l^{2}J^{2}(r_{+}^{2}+r_{-}^{2})-l^{4}J^{4}}{16l^{4}r_{+}^{3}r_{-}^{3}},
\end{equation}
The   entropy and temperature of this black hole can be used to write the Komar energy of the black hole,  \cite{Komar},
\begin{eqnarray}\label{8}
E_{0\pm}&=&2S_{0\pm}T_{\pm}=\frac{4r_{\pm}^{4}-l^{2}J^{2}}{l^{2}r_{\pm}^{2}}v_{f}^{3}=4v_{f}^{3}\left[M-\frac{2J^{2}}{r_{\pm}^{2}}\right]\nonumber\\
&=&4lv_{f}^{3}\frac{(lM)^2\pm lM\sqrt{(lM)^{2}-J^{2}}-J^{2}}{lM\pm\sqrt{(lM)^{2}-J^{2}}}.
\end{eqnarray}
Again, the effects of thermal fluctuations have been neglected. The product of the Komar energies  can be expressed as
\begin{equation}\label{9}
E_{0+}E_{0-}=16v_{f}^{6}\left[M^{2}-\frac{2MJ^{2}}{r_{+}^{2}}-\frac{2MJ^{2}}{r_{-}^{2}}+\frac{4J^{4}}{r_{+}^{2}r_{-}^{2}}\right].
\end{equation}
It may be noted that the product of  the temperatures and Komar energies depends on the black hole mass, while $A_{+}A_{-}$ and $S_{0+}S_{0-}$ are independent of the mass of the black hole.

Using the Eq. (\ref{5}),  we can obtain ADM mass of the black hole
\begin{equation}\label{10}
M_{0ADM}=\frac{S_{0\pm}^{2}}{16\pi^{2}l^{2}}+\frac{4\pi^{2}J^{2}}{S_{0\pm}^{2}}.
\end{equation}
Now, the first law of thermodynamics for this black hole may be
\begin{equation}\label{11}
dM_{0ADM}=T_{\pm}dS_{0\pm}+\Omega_{0\pm} dJ,
\end{equation}
where
\begin{equation}\label{12}
\Omega_{0\pm}=\left(\frac{dM_{0ADM}}{dJ}\right)_{S_{0\pm}}=\frac{8\pi^{2}J}{S_{0\pm}^{2}}.
\end{equation}
and
\begin{equation}\label{13}
T_{\pm}=\left(\frac{dM_{0ADM}}{dS_{0\pm}}\right)_{J}.
\end{equation}
It may be noted that the  first law of thermodynamics satisfied for this black hole solution.
We can also calculate the  specific heat of this black hole solution as
\begin{equation}\label{14}
C_{0\pm}=\frac{dM_{0ADM}}{dT_{\pm}}=4\pi v_{f}^{3} r_{\pm}\frac{4r_{\pm}^{4}-l^{2}J^{2}}{4r_{\pm}^{4}+3l^{2}J^{2}}.
\end{equation}
Thus, this black hole is in stable phase ($C_{0+}\geq0$) for $lM\geq J$, and $C_{0-}<0$  for $lM\geq J$.
\section{Corrected thermodynamics}
\begin{figure}[h!]
 \begin{center}$
 \begin{array}{cccc}
\includegraphics[width=60 mm]{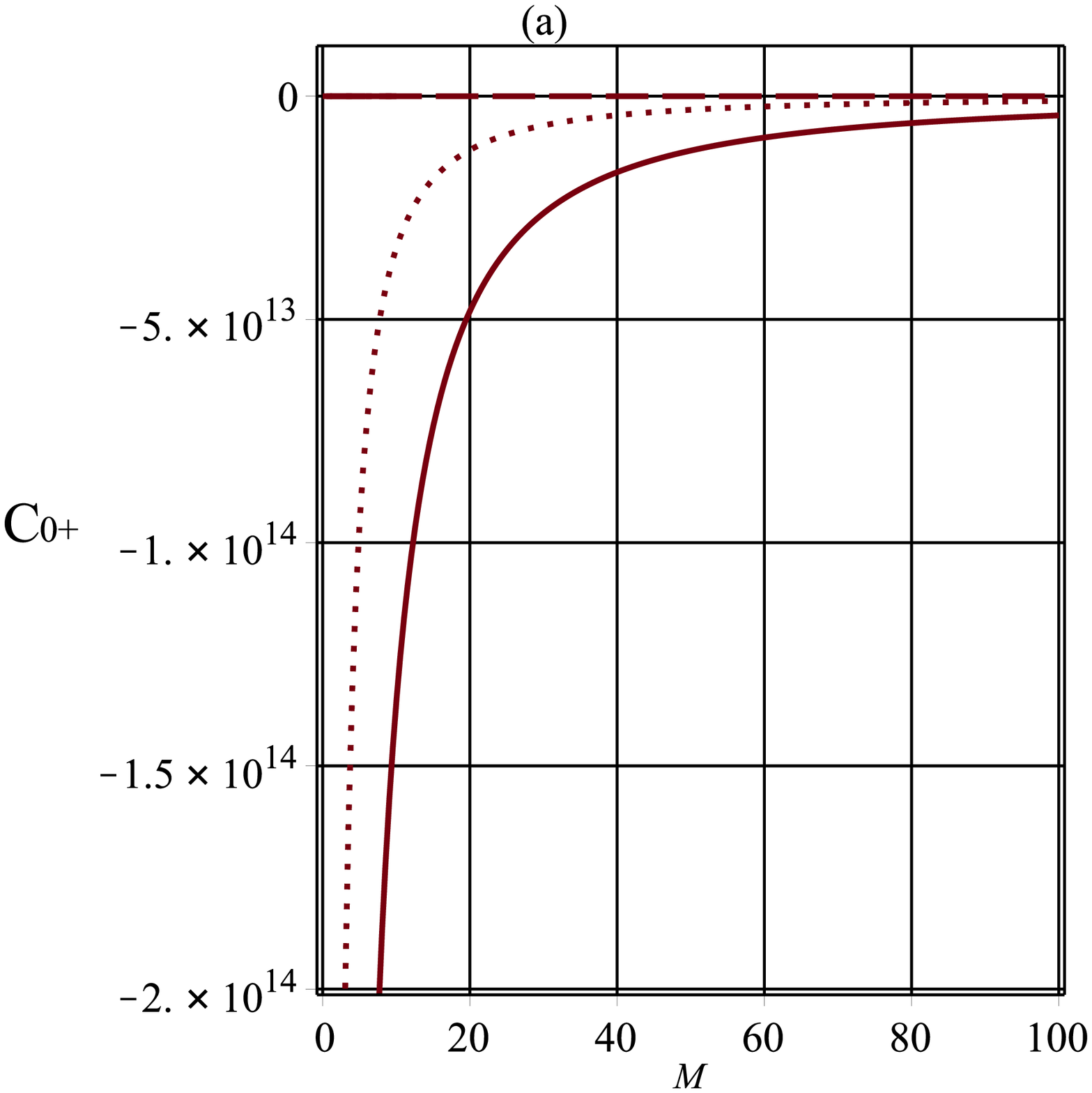}&\includegraphics[width=60 mm]{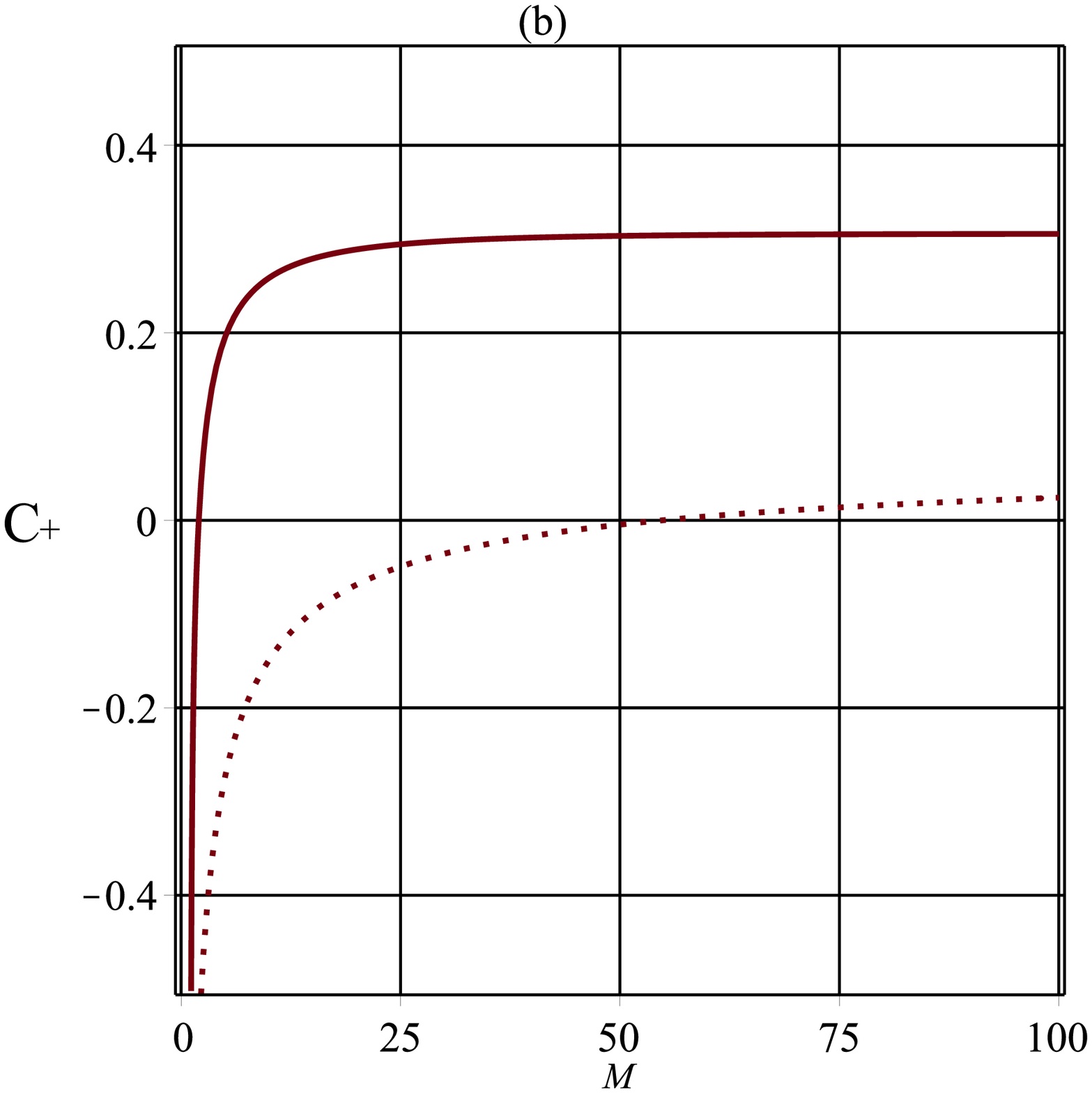}\\
\includegraphics[width=60 mm]{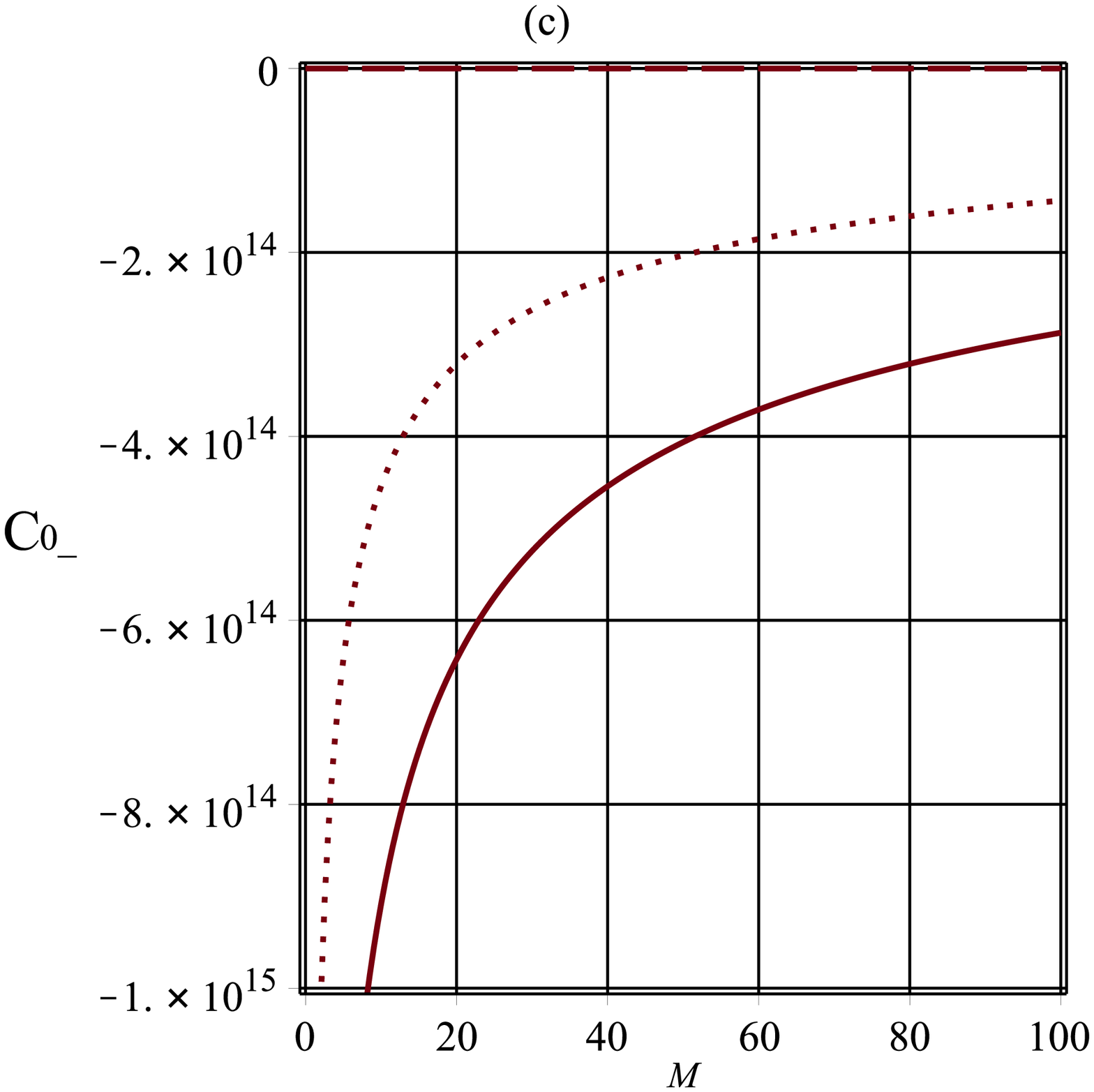}&\includegraphics[width=60 mm]{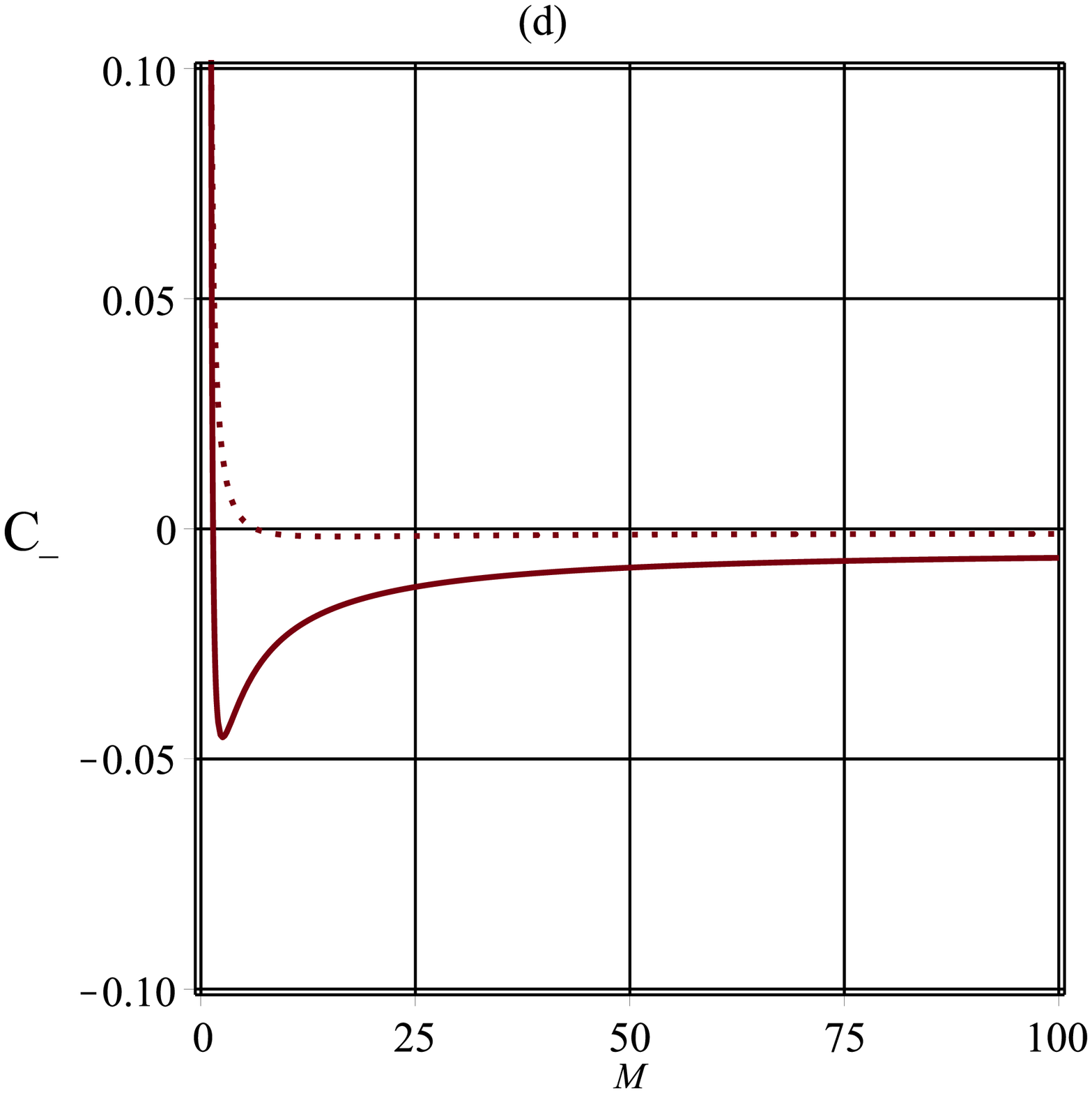}
 \end{array}$
 \end{center}
\caption{Specific heat in terms of the black hole mass with $l=1$ and $v_{f}=0.003$. (a) $\alpha=0$, $J=0$ dashed line, $J=0.5$ dotted
line, $J=1$ solid line. (b) $\alpha=1$, $J=0.5$ dotted line, $J=1$ solid line. (c) $\alpha=0$, $J=0$ dashed line, $J=0.5$
dotted line, $J=1$ solid line. (d) $\alpha=1$, $J=0.5$ dotted line, $J=1$ solid line.}
 \label{fig1}
\end{figure}

\begin{figure}[h!]
 \begin{center}$
 \begin{array}{cccc}
\includegraphics[width=52 mm]{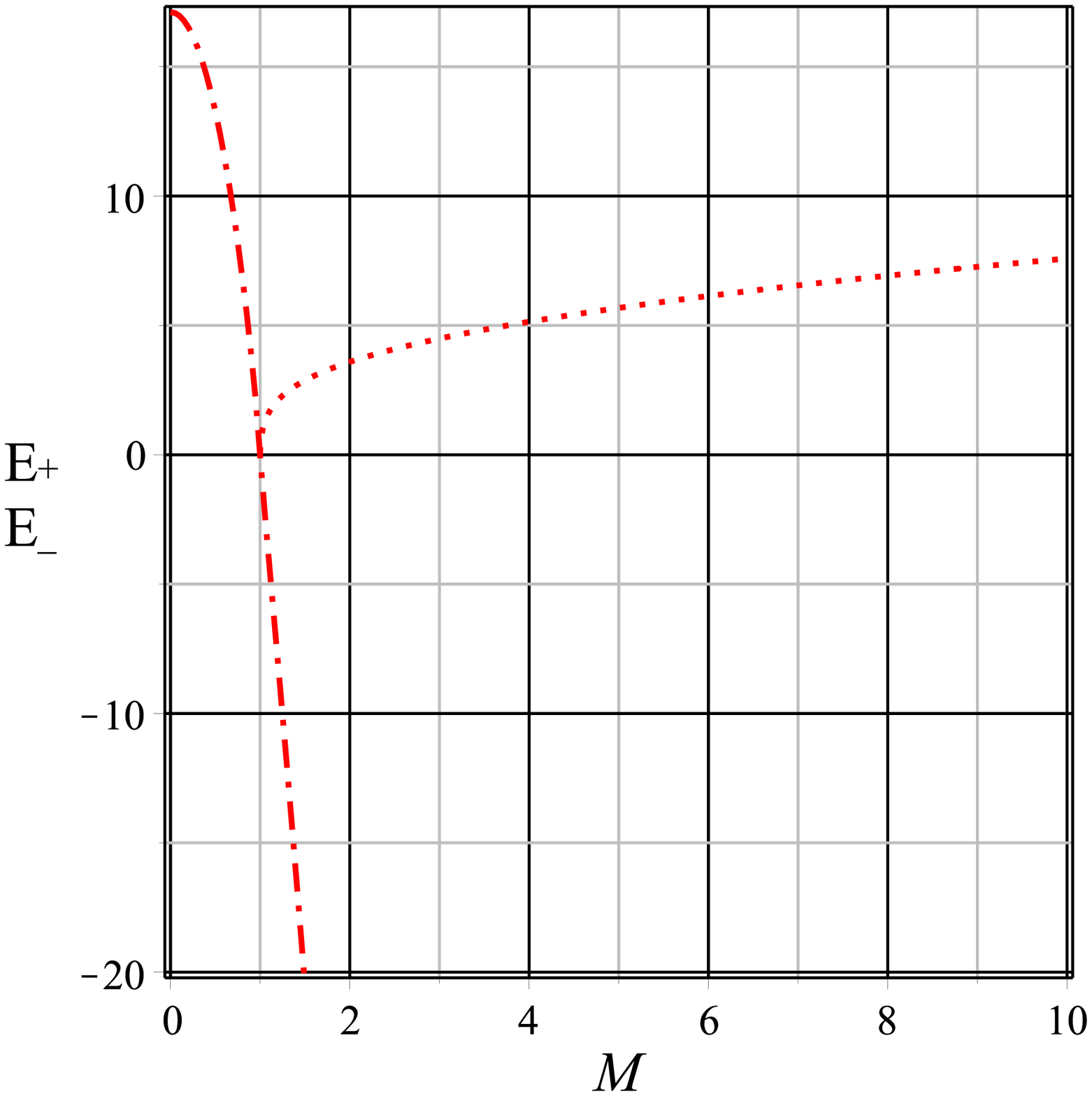}&\includegraphics[width=52 mm]{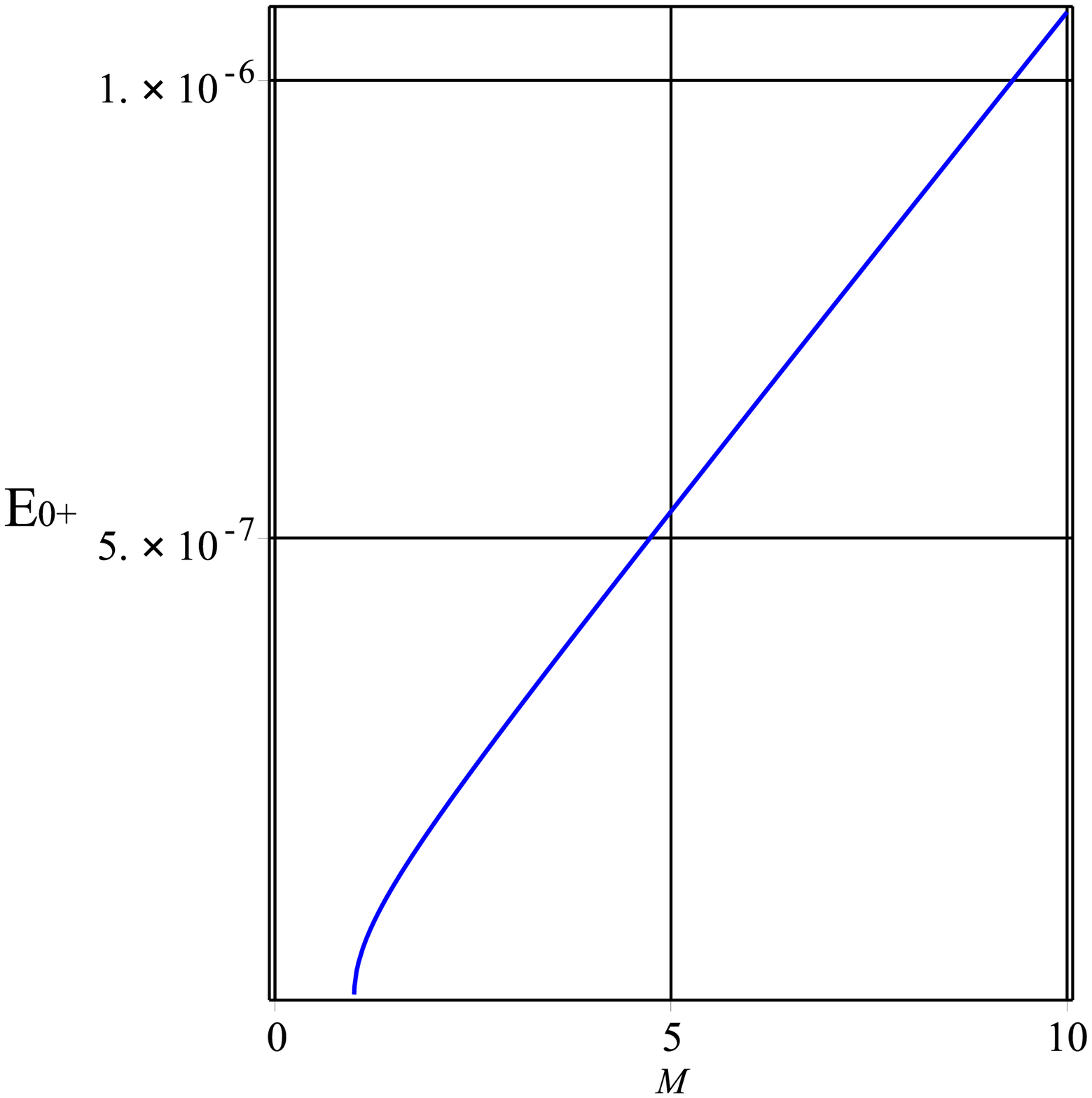}&\includegraphics[width=52 mm]{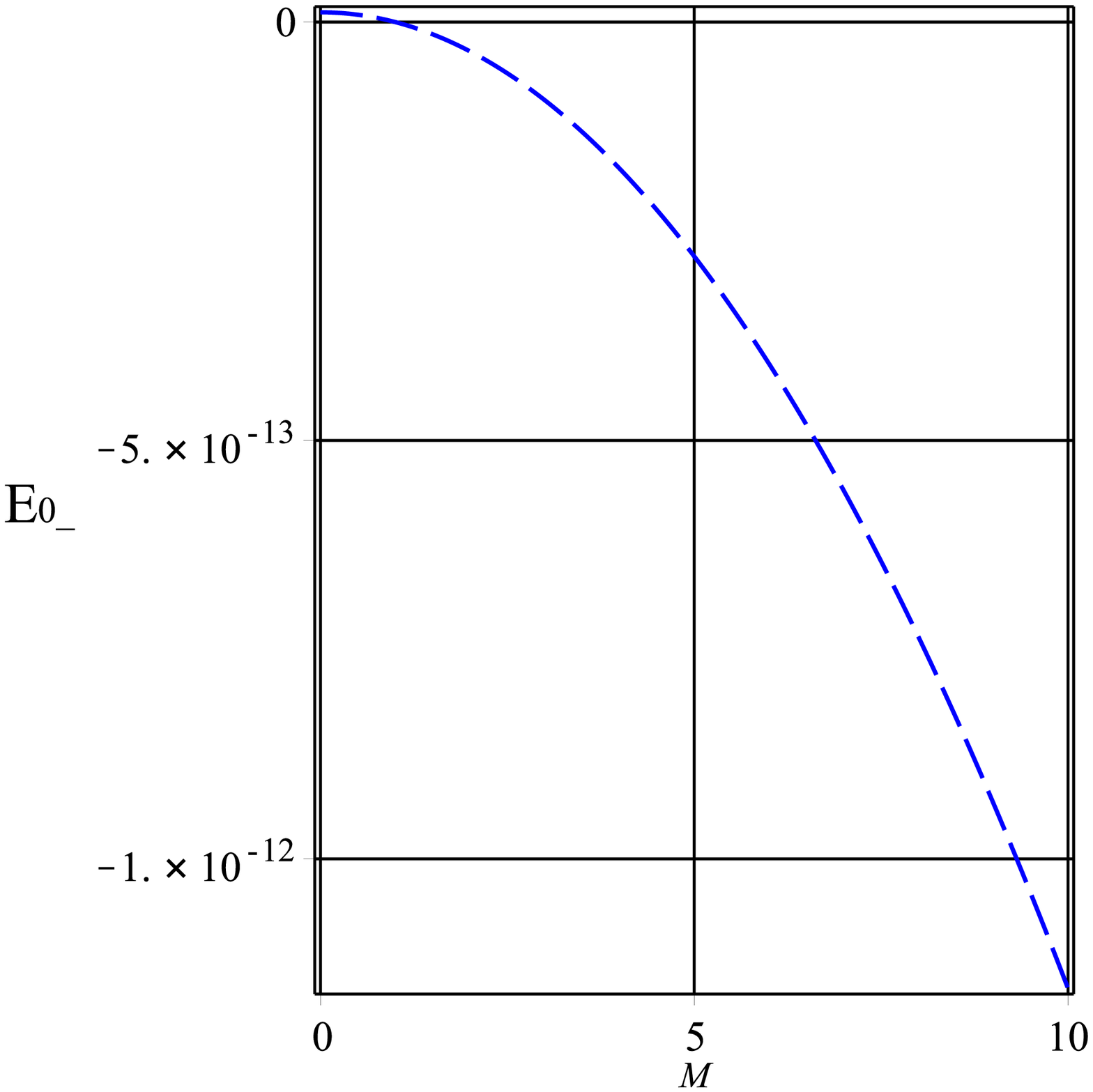}
 \end{array}$
 \end{center}
\caption{Energy in terms of the black hole mass with $l=1$ and $J=1$. $E_{0-}$ denoted by Blue dashed line, while $E_{0+}$
denoted by Blue solid line. $E_{-}$ denoted by Red dash dotted line, while $E_{+}$ denoted by Red dotted line.}
 \label{fig2}
\end{figure}

We can now include the  thermal fluctuations and analyze the produced effects by them.  Thus, we will write the corrected entropy of this
black hole as
\begin{equation}\label{15}
S_{\pm}=S_{0\pm}-\frac{\alpha}{2}\ln{S_{0\pm}T_{\pm}^{2}},
\end{equation}
We expected that $C_{\pm}\geq0$  for $lM\geq J$, i.e.,  both $C_{+}$ and $C_{-}$ have positive values.
In the Fig. \ref{fig1},  we have plotted the  value of the specific heat. It may be noted that for  $C_{-}$ to be positive, the
logarithmic correction has to be included. In Fig. $1 (a)$, it is demonstrated that $C_{+}$ without logarithmic correction is negative for any choice of $J$ and $M$. In the presence of logarithmic correction, we have $C_{+}\geq0$ for the massive black hole. Thus, for $C_{\pm}\geq0$,
with appropriate choice of the black hole mass, the effect of thermal fluctuations has to be considered.
In that case,  we have $E_{\pm}=2S_{\pm}T_{\pm}$. It is easy to show that $E_{+}\leq E_{0+}$, and the equality hold for small masses.
 We also observe that $E_{0-}$, as well as $E_{-}$, may change the sign for a massive black hole. It is clear that logarithmic correction increases the value of the energy. This situation
is illustrated by the plots of the Fig. \ref{fig2}. Furthermore,  $E_{0+}E_{0-}$ is negative for a massive black hole, and $E_{+}E_{-}$
is positive for a massive black hole. For the small values $M$,  ($Ml\approx2J$),  we have $E_{0+}E_{0-}=E_{+}E_{-}$.\\

\begin{figure}[h!]
 \begin{center}$
 \begin{array}{cccc}
\includegraphics[width=50 mm]{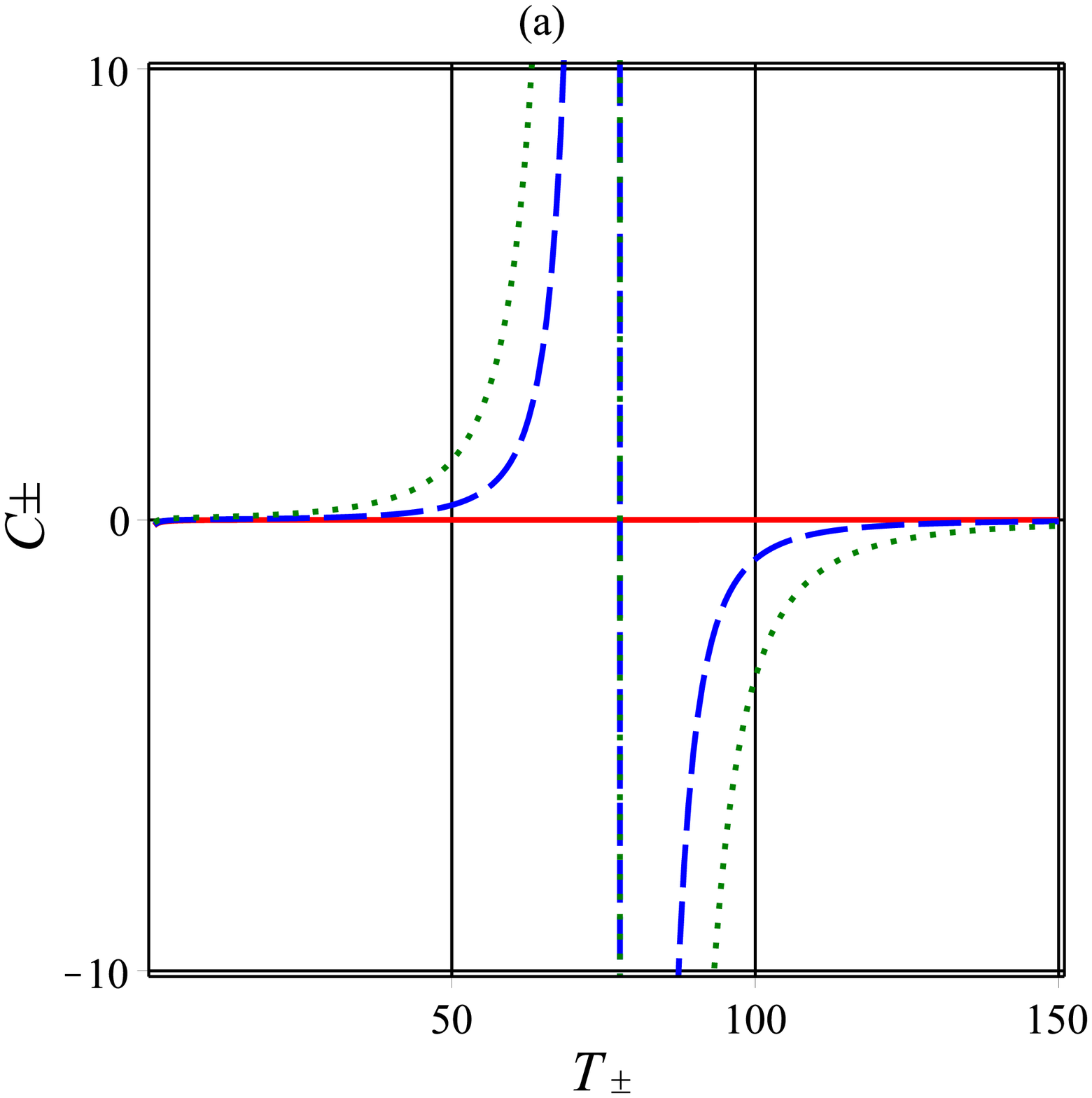}\includegraphics[width=50 mm]{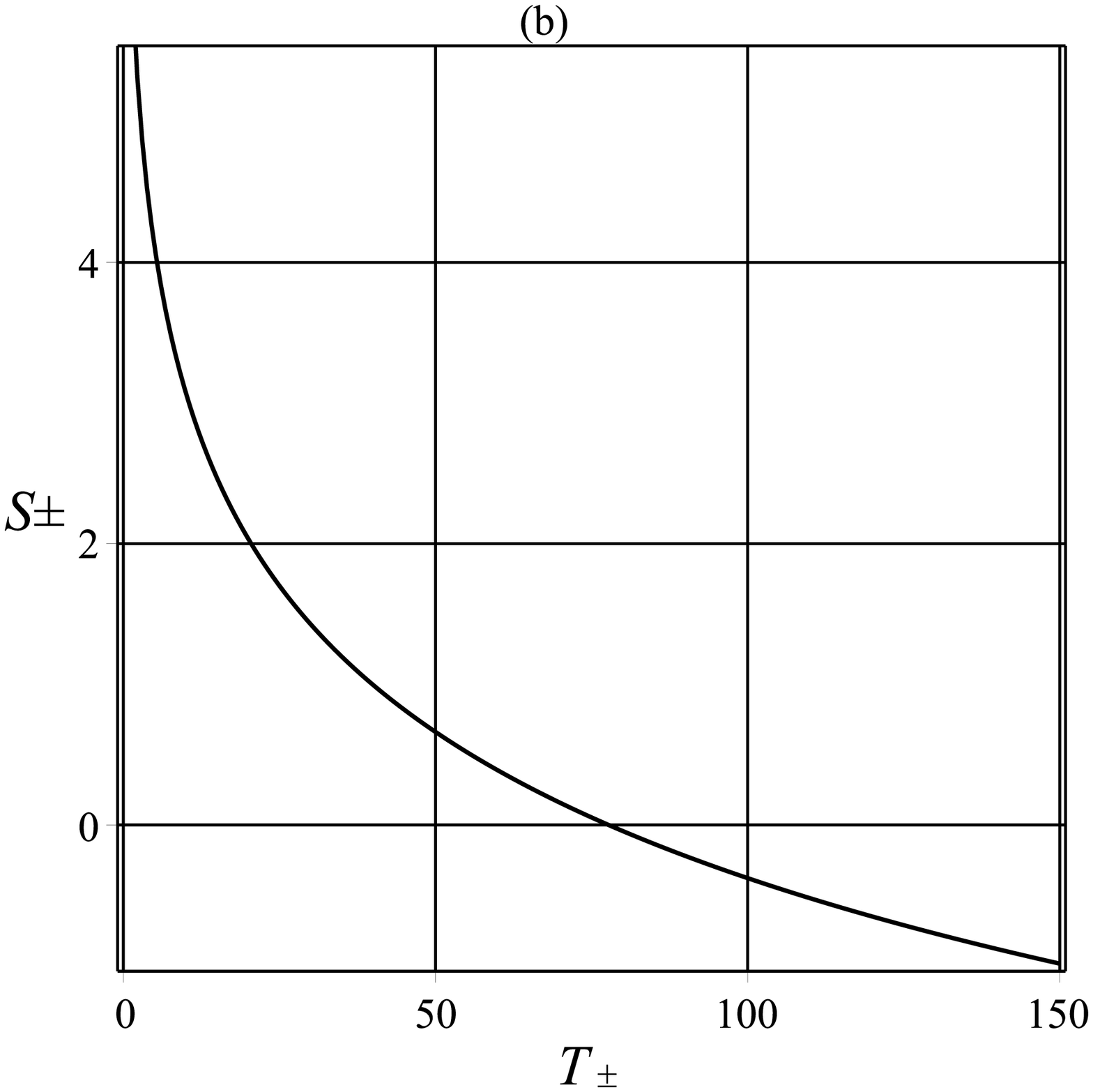}\includegraphics[width=50 mm]{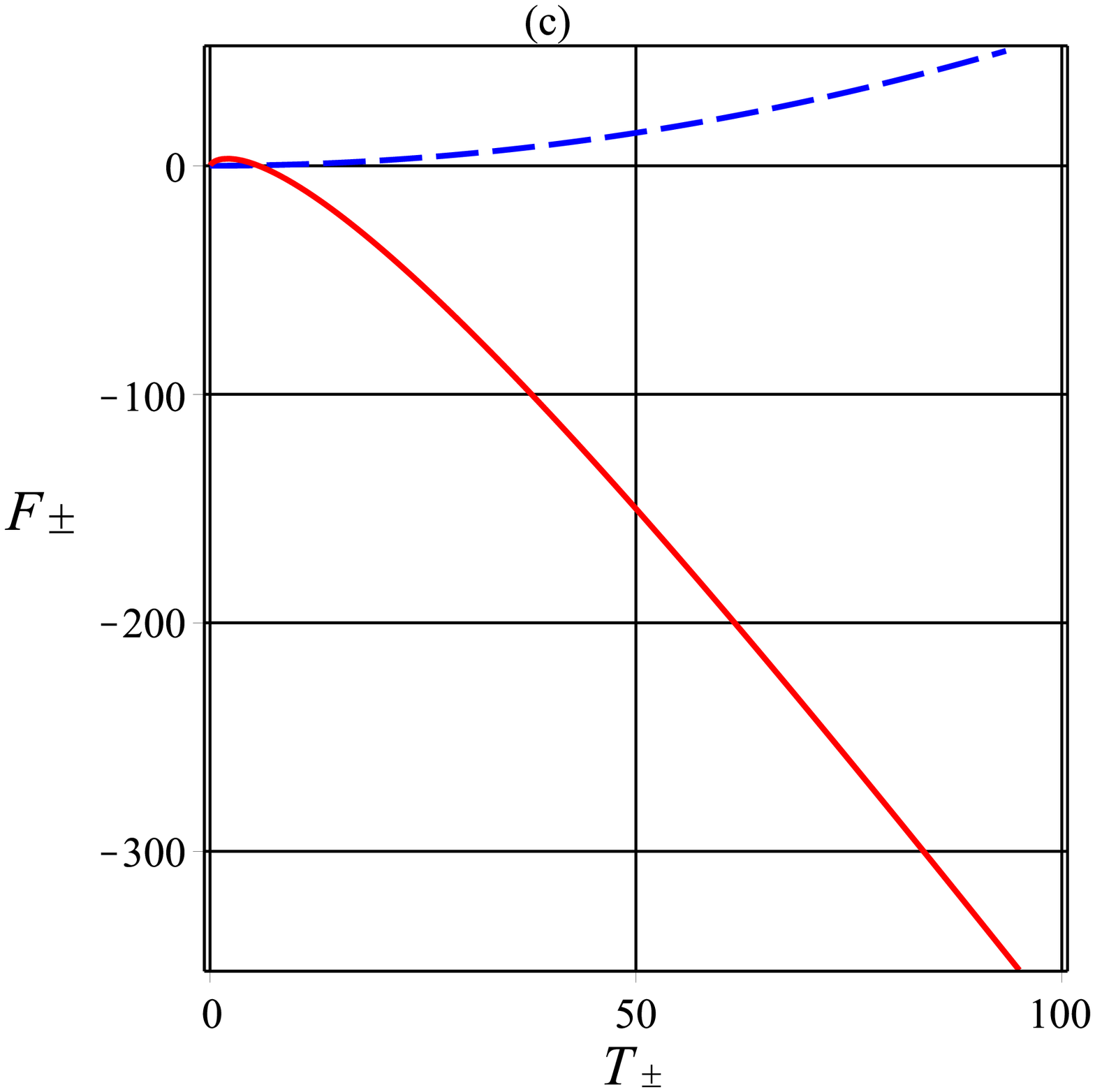}
 \end{array}$
 \end{center}
\caption{(a) Specific heat in term of temperature with $l=1$, $v_{f}=0.003$, $\alpha=1$. $J=0$ (solid red), $J=0.2$ (dashed blue), $J=0.4$ (dotted green). (b) Entropy in term of temperature with $l=1$, $v_{f}=0.003$, $\alpha=1$. (c) Helmholtz free energy in terms of temperature with $l=10$, $v_{f}=0.009$, and $J=0.1$; $\alpha=0$ (dashed blue), $\alpha=1$ (solid red).}
 \label{fig5}
\end{figure}

To have a comparison with the adaptive model of graphene, we should represent our results in terms of temperature. In the case of $lJ\ll1$ one can obtain analytical expressions. In the Fig. \ref{fig5} (a) we can see temperature dependence specific heat. As before, in the case of $\alpha=0$ we find negative specific heat, but in the presence of logarithmic correction we see a phase transition. Specific heat is positive for the low temperature, while it is negative for the higher temperature. It means that at the higher temperature, graphene structure is broken and entropy will be negative (as illustrated by the Fig. \ref{fig5} (b)). Asymptotic behavior of specific heat is corresponding to maximum value of the specific heat of the adaptive model of graphene \cite{39}. We find that value of the corrected entropy does not affected by small $J$. Also, in the Fig. \ref{fig5} (c) we can see behavior of Helmholtz free energy in terms of temperature and see that is increasing function of temperature with positive value for the case of $\alpha=0$ while there is a maximum for the logarithmic corrected case at very low temperature, and general behavior of Helmholtz free energy is increasing function of the temperature with negative value, which is in agreement with behavior of the adaptive model of graphene \cite{39}.

\section{Hawking radiation}
\begin{figure}[h!]
 \begin{center}$
 \begin{array}{cccc}
\includegraphics[width=75 mm]{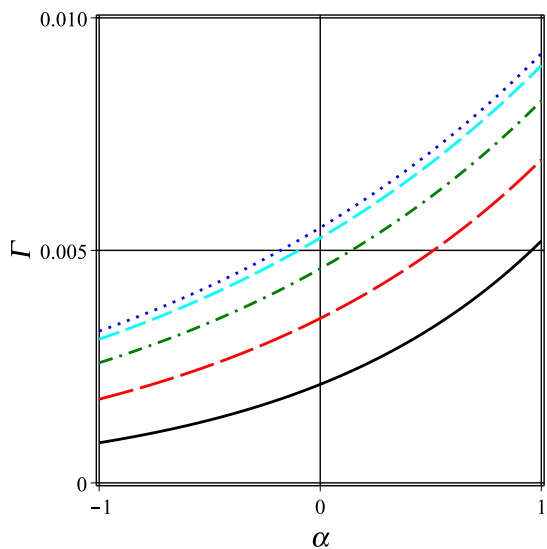}
 \end{array}$
 \end{center}
\caption{Emission rate of Hawking radiation in terms
of the logarithmic correction parameter with $l=1$, $\omega=1$ and $M=2$.
$J=0$ (blue dot), $J=0.2$ (cyan dash), $J=0.4$ (green dash dot), $J=0.6$
(red long dash), $J=0.8$ (black solid).}
 \label{fig3}
\end{figure}

Emission rate of Hawking radiation given by \cite{end},
\begin{equation}\label{E1}
\Gamma=e^{\Delta S_{+}},
\end{equation}
which obtained from the action for an outgoing positive-energy particle which
crosses the event horizon outwards from
\begin{equation}\label{E2}
r_{in+}=\sqrt{\frac{l}{2}(lM+\sqrt{(lM)^{2}-J^{2}})},
\end{equation}
to
\begin{equation}\label{E3}
r_{out+}=\sqrt{\frac{l}{2}(l(M-\omega)+\sqrt{(l(M-\omega))^{2}-J^{2}})},
\end{equation}
so $\Delta S_{+}=S(r_{out+})-S(r_{in+})$. In the Fig. \ref{fig3} we can see the effect of thermal fluctuations on the emission rate of Hawking radiation given by the equation (\ref{E1}). It is clear that thermal fluctuation increases the value of the emission rate.\\

\begin{figure}[h!]
 \begin{center}$
 \begin{array}{cccc}
\includegraphics[width=75 mm]{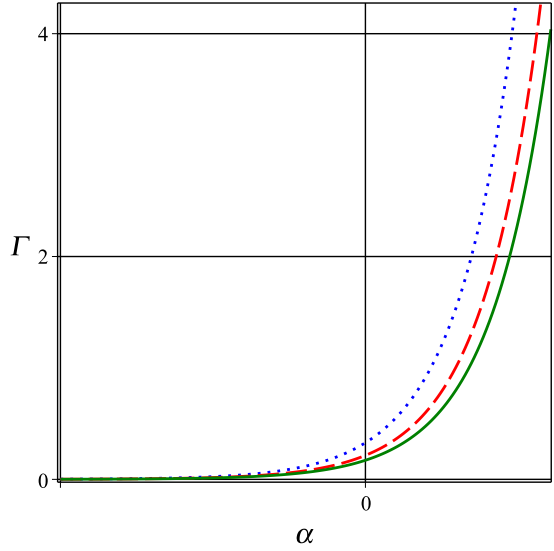}
 \end{array}$
 \end{center}
\caption{Emission rate of Hawking radiation in terms of the
logarithmic correction parameter with $J=0$, $\omega=0.5$ and $M=1$. $c_{3}=0.5$
(blue dot), $c_{3}=1$ (red dash), $c_{3}=2$ (green solid).}
 \label{fig4}
\end{figure}

We can use $J=0$ limit of BTZ black hole with $(-,+,+,+)$ signature as ansatz for graphene \cite{IJMPD}, hence rewrite metric (\ref{1}) as follow,
\begin{equation}\label{BTZJ0}
ds_{BTZ}^{2}(J=0)=-\left(\frac{{\mathcal{R}}^{2}}{c^{2}}-M\right)dt^{2}+\frac{d{\mathcal{R}}^{2}}{\frac{{\mathcal{R}}^{2}}{c^{2}}-M}-{\mathcal{R}}^{2}d\chi^{2},
\end{equation}
where $c$ and $M$ are positive constants, and $\chi$ is the angular variable ($\phi$). It can be rewrite as,
\begin{equation}\label{BTZJ0-1}
ds_{BTZ}^{2}(J=0)=-\left(\frac{r^{2}}{c^{2}}-M\right)dt^{2}+\frac{c^{2}r^{2}-Mc^{4}}{(r^{2}-r_{+}^{2})^{2}}dr^{2}+\frac{(r^{2}-Mc^{2})r^{2}}{r^{2}-r_{+}^{2}}
d\chi^{2},
\end{equation}
where we used our notation for radial coordinate ${\mathcal{R}}\equiv r$. Also, $r{+}=c\sqrt{M}$ is the black hole event horizon. In that case, the black hole temperature can be written as,
\begin{equation}\label{TBTZJ0-1}
T=c_{1}r_{+},
\end{equation}
where $c_{1}=(2\pi c^{2})^{-1}$ is a positive constant. Also, the black hole entropy given by,
\begin{equation}\label{S0BTZJ0-1}
S_{0}=c_{2}r_{+},
\end{equation}
where $c_{2}=4\pi c^{3}$ is a positive constant. Hence, we can write,
\begin{equation}\label{S0TBTZJ0-1}
S_{0}=c_{3}T,
\end{equation}
where $c_{3}=8\pi^{2} c^{5}$ is another positive constant. Of course, the ansatz for graphene tell $c=l$. Therefore, we have the following corrected entropy,
\begin{equation}\label{STBTZJ0-1}
S=c_{3}T-\frac{\alpha}{2}\ln{(c_{3}T^{3})}.
\end{equation}
In that case, the specific heat obtained as,
\begin{equation}\label{CTBTZJ0-1}
C=T\left(\frac{dS}{dT}\right)=c_{3}T-\frac{3\alpha}{2}.
\end{equation}
It is clear that the effect of logarithmic correction is a reduction of specific heat (for positive $\alpha$ as usual). Now we can obtain internal energy as,
\begin{equation}\label{UTBTZJ0-1}
U=\frac{c_{3}}{2}T^{2}-\frac{3\alpha}{2}T.
\end{equation}
We can see that the logarithmic correction decreased the value of the internal energy. By using the entropy (\ref{STBTZJ0-1}) and internal energy (\ref{UTBTZJ0-1}) we can obtain Helmholtz free energy as follow,
\begin{equation}\label{FTBTZJ0-1}
F=E-TS=\frac{c_{3}}{2}T^{2}-(c_{3}+\frac{3\alpha}{2})T+\frac{\alpha}{2}\ln{(c_{3}T^{3})}.
\end{equation}
In general, we can say that effect of logarithmic correction is decreasing of thermodynamics potentials like Helmholtz free energy.
Without logarithmic correction ($\alpha\neq0$) there is a minimum for the free energy while in the presence of logarithmic correction there
is no minimum point for the Helmholtz free energy instead there is an inflection point.\\
As before we can investigate emission rate of Hawking radiation which illustrated by the Fig. \ref{fig4}.

\section{Randers optical metric}
The BTZ like metric (\ref{1}) can conformally transform to the Randers optical metric,  and in a sheet of graphene, it can be written as \cite{1}
\begin{equation}\label{R1}
ds_{R}^{2}=\frac{dr^{2}}{\Delta\Delta_{0}}+\frac{1}{\Delta_{0}^{2}}\left(\Delta r^{2}-\frac{J^{2}}{4}\right)d\phi^{2}-v_{f}^{2}dt^{2}-\frac{J}{\Delta_{0}}v_{f}dtd\phi,
\end{equation}
where $
\Delta_{0}=\frac{r^{2}}{l^{2}}-M.
$,  and the value of $\Delta$ is similar to its value in the original BTZ metric.
So, the value of the original entropy,  $S_{0\pm}$,  is also similar to the
 the value obtained in Eq. (\ref{5}). The Hawking temperature of the inner horizon and the outer horizon is now  given by,
\begin{eqnarray}\label{R3}
T_{\pm}&=&\frac{8r_{\pm}^{6}-8Ml^{2}r_{\pm}^{4}+Ml^{4}J^{2}}{8\pi l^{4} r_{\pm}^{3}}\nonumber\\
&=&\frac{\sqrt{2}J^{2}}{4\pi l^{\frac{5}{2}}} \frac{J^{2}-(lM)^2\pm lM\sqrt{(lM)^{2}-J^{2}}}{(lM\pm\sqrt{(lM)^{2}-J^{2}})^{\frac{5}{2}}}.
\end{eqnarray}
The product of temperatures on both the horizons can now be written as
\begin{eqnarray}\label{R4}
4\pi^{2} T_{+}T_{-}&=&\frac{64Ml^{2}r_{+}^{4}r_{-}^{4}(Ml^{2}-r_{-}^{2}-r_{+}^{2}+\frac{r_{-}^{2}r_{+}^{2}}{Ml^{2}})}{16l^{4}r_{+}^{3}r_{-}^{3}}\nonumber\\
&+&\frac{8MJ^{2}l^{4}(r_{+}^{6}+r_{-}^{6}-Ml^{2}(r_{+}^{4}+r_{-}^{4}))+J^{4}M^{2}l^{8}}{16l^{4}r_{+}^{3}r_{-}^{3}}.
\end{eqnarray}
Now, we can again analyze the effect of thermal fluctuations on the thermodynamics of the black holes.
We observe that  the energy and specific heat are negative even in the presence of
thermal fluctuations.  The  Zermelo optical metric  in a sheet of graphene  can be written as   \cite{1}
\begin{equation}\label{R5}
ds_{Z}^{2}=\frac{dr^{2}}{\Delta^{2}}+\frac{r^{2}}{\Delta}d\phi^{2}+(\frac{J^{2}}{4\Delta r^{2}}-1)v_{f}^{2}dt^{2}-\frac{J}{\Delta}v_{f}dtd\phi.
\end{equation}
Here the temperature on the horizon is given by  $T_{\pm}=0$,
so specific heat is zero, and in absence of thermal fluctuations the  entropy is given by the Eq.  (\ref{5}). Hence, the   entropy in presence of thermal
fluctuations will
be infinite and system will remain  in equilibrium. Thus, using Eq. (\ref{13}), we obtain,
\begin{equation}\label{R6}
\left(\frac{dM_{0ADM}}{dS_{0\pm}}\right)_{J}=\frac{S_{0\pm}}{8\pi^{2}l^{2}}-\frac{8\pi^{2}J^{2}}{S_{0\pm}^{3}}=0.
\end{equation}
So, in this case we have
$
lM=J,$
and a black hole is in the extremal limit where $r_{+}=r_{-}$.
Thus, we can analyze the effect of logarithmic correction of a sheet of graphene. This can be used to obtain the correct value of $\alpha$ comparing the adaptive model of graphene. As even the real black holes can be considered as thermodynamical objects
in the Jacobson formalism \cite{z12j,jz12}, we will argue that this is the value of $\alpha$ which should occur even in real
black holes. Thus, we can select the correct approach to quantum gravity.

\section{Summary}
We can summarize our argument as follows:
{
\enumerate
 \item It may be noted that in the effective field theory describing graphene,
the velocity of light $c$ gets replaced by a much smaller Fermi velocity $v_f$. However, the main symmetries of background geometry
describing a flat sheet of graphene is an effective $(2+1)d$ Lorentz symmetry, and any curvature in the sheet of graphene is described
as a curved $(2+1)d$ space-time. Thus, it is possible to have an effective horizon in graphene, and in this effective horizon the
Fermi velocity effectively acts like the velocity of light for such analogous black holes.
\item
It is known that  almost all approaches to  quantum gravity  predict the logarithmic corrections to the entropy of a black hole,
but the coefficient
of this logarithmic term differs between different approaches to quantum gravity. Hence, this coefficient
can be used to select between the different approaches to quantum gravity. So, if we can know what the correct coefficient
for the logarithmic term, we can select the correct approach to quantum gravity.
\item
Analogous black holes can be used to  test the correct coefficient for the logarithmic term,
as it has been demonstrated that
an analogous BTZ like metric can exist in graphene \cite{1, u}. So, in this paper,
we analyzed the effects of the logarithmic corrections
to the entropy of such an analogous black hole  in adaptive model of graphene. We keep the coefficient
of this BTZ like black hole as a variable.
Thus, in this paper, we analyze the effects of the logarithmic corrections to the entropy of a
BTZ like black holes in adaptive model of graphene,
with a variable coefficient for this logarithmic term.

\item
It is possible to calculate the effective Hawking-like radiation in graphene, and also analyze the effect of logarithmic corrections on
such effective processes. However, such effective Hawking radiation can be analyzed in experimentally using curved sheet of graphene,
and thus, the coefficients of such a term can be fixed. Thus, we can use the adaptive model of graphene to
fix the coefficient of quantum correction
to an analogous effective black hole.
\item
It may be noted that there is no fundamental difference between analogous black holes and real
black holes in the Jacobson formalism \cite{z12j,jz12}.
As in this formalism, even the real black holes (and all of geometry in general relativity)
are some thermodynamical objects, just like the  analogous black hole. However, we still expect
logarithmic corrections even in Jacobson formalism and analogous black holes due to
thermal fluctuations \cite{l1,SPR}.

}
\section{Conclusion and discussion}
In this paper, we analyzed the black hole like solution that has been that occurs in the effective
field theory describing  an adaptive model of graphene. Although it is still far from real graphene but it is possible to study from several points of view like superconductivity \cite{Sa1, Sa2}.\\
It is possible to obtain an effective BTZ-like solution in graphene
\cite{1, u}. Here the velocity of light gets replaced by the Fermi velocity. However, it is possible to have a horizon
for such an analogous black hole, and just as a black hole traps photon moving with the velocity of light, this
analogous black hole traps fermions moving with the Fermi velocity.
Furthermore, the Hawking radiation from   analogous back holes in graphene has been studied \cite{1, u, u1}, and
so, we analyze the entropy associated with such back holes.
In fact, we analyze the corrections to the entropy of such black holes because of thermal fluctuations,
and keep the coefficient of such a correction term as a variable. This is because almost all approaches to quantum gravity
predict that the corrections to the entropy of a black hole are a logarithmic correction term, but the coefficient
of such a term varies between different approaches to quantum gravity. So, if we can experimentally verify the correct
coefficient to the logarithmic correction term, we can select the correct approach to quantum gravity. As
we cannot test the black hole thermodynamics directly, in this work, we proposed to analyze it using
an analogous black hole like solution in the adaptive model of graphene.
In fact, there is no fundamental difference between
analogous  black holes and real
black holes in the Jacobson formalism \cite{z12j,jz12}, except that the velocity of light gets replaced by the Fermi velocity.
This  even real black holes can be described as thermodynamical objects in the
Jacobson formalism, and the effect of quantum fluctuation can be described in terms of
thermal fluctuations \cite{l1,SPR}. We should notice that the logarithmic corrected entropy used in this paper has similar shape as the entropy studied in a $Co-C$ system \cite{theory}.
In this system the  low entropy in graphene was studied using  the $Co-C$ system   with activation energy values of order $0.1-1 eV$,
which coincident with
the internal energy is given by the Fig. \ref{fig2} in the presence of logarithmic correction for small  black hole masses.
The  specific heat is of order unity \cite{thermal}, which suggest presence of thermal correction (see Fig. \ref{fig1}).
Without thermal correction, specific heat is of order $10^{14}$ with negative value while we find specific heat is about $0.3$ in presence of
logarithmic correction, and so it is close to experimental results.
It is also interesting to compute thermal conductivity of this system and compare it with experimental results \cite{exp}.\\
Now,  it is possible to test the effect of logarithmic correct on the effective
black hole like solution in the adaptive model of graphene, and we have explicitly analyzed its effect on the effective Hawking radiation
from such analogous black holes. These results may be tested experimentally and used to determine the correct coefficient
for the logarithmic correction term to the entropy of a black hole. This can in turn be used to select the correct
approach to quantum gravity, which would produce such a coefficient for the logarithmic correct term of a black hole.
Thus, it is possible to analyze quantum gravitational effects using analogous black hole like solution in graphene.

\section{Acknowledgments}
We would like to thanks Alfredo Iorio for useful suggestions that helped us improve this paper.

\end{document}